\documentclass{elsart3p}
\usepackage{natbib}
\usepackage{amssymb}
\usepackage{graphicx}
\journal{New Astronomy Review}
\begin{document}
\begin{frontmatter}
\title{Radio-loudness of Active Galaxies and the Black Hole Evolution}
\author[label1]{Marek Sikora}
\ead{sikora@camk.edu.pl}
\author[label2,label3]{{\L}ukasz Stawarz}
\author[label4,label3]{Jean-Pierre Lasota}
\address[label1]{Nicolaus Copernicus Astronomical Center,
ul. Bartycka 18, 00-716 Warszawa, Poland}
\address[label2]{Kavli Institute for Particle Astrophysics and Cosmology,
Stanford University, Stanford CA 94305}
\address[label3]{Astronomical Observatory, Jagiellonian University,
ul. Orla 171, 30-244 Krak\'ow, Poland}
\address[label4]{Institut d'Astrophysique de Paris, UMR 7095 CNRS,
UPMC Univ Paris 06, 98bis Bd Arago, 75014 Paris, France}

\begin{abstract}
Active galactic nuclei (AGNs) form two distinct sequences on the
radio-loudness -- Eddington-ratio plane. The `upper' sequence
contains radio selected AGNs, the `lower' sequence is composed
mainly of optically selected AGNs. The sequences mark the upper
bounds for the radio-loudness of two distinct populations of
AGNs, hosted respectively by elliptical and disk galaxies. Both
sequences show the same dependence of the radio-loudness on the
Eddington ratio (an increase with decreasing Eddington ratio), which
suggests that another parameter in addition to the accretion rate
must play a role in determining the efficiency of jet production in
AGNs. We speculate that this additional parameter is the spin
of the black hole, assuming that black holes in giant elliptical
galaxies have (on average) much larger spins than black holes in
disc galaxies. Possible evolutionary scenarios leading to such a
spin dichotomy are discussed. The galaxy-morphology related
radio-dichotomy breaks down at high accretion rates where the
dominant fraction of luminous quasars being hosted by giant
ellipticals is radio quiet. This indicates that the production
of powerful jets at high accretion rates is in most cases suppressed
and, in analogy to X-ray binary systems (XRB) during high and
very high states, may be intermittent. Such intermittency can be
caused by switches between two different accretion modes,
assuming that only during one of them an outflow from the central
engine is sufficiently collimated to form a relativistic jet.
\end{abstract}


\end{frontmatter}

\section{Introduction}

Already in the early 1960s it became clear that the
strongest radio sources associated with disk galaxies are by about 3
orders of magnitude weaker than the strongest radio sources
hosted by giant elliptical galaxies, and that the
most radio luminous objects in the Universe are quasars
\citep{mms64}. Soon it was realized, however, that the majority of
quasars is radio quiet
\citep{Sandage,shp80}. Initial optical imaging indicated 
that their
radio-bimodality may be related to the host galaxy
morphology, 
with radio quiet quasars preferentially hosted by disk galaxies and radio-loud 
quasars hosted by elliptical galaxies \citep{Mal84,shbrb86}.
This suggested a possible association of radio quiet-quasars with 
Seyfert galaxies and radio-loud quasars with radio galaxies.
Such a bimodality     
is visualized by \citet{xlb99} in the radio-luminosity
versus [O{\sc III}]-luminosity representation.

Since radio structures in AGNs are powered by jets, the AGN radio
dichotomy is most likely related to the efficiency of a jet
production. According to the spin paradigm, this efficiency
depends on the value of the black hole spin \citep{bz77,bland90}.
In the simplest version of this paradigm the
observed radio dichotomy is directly related to
the cosmologically determined distribution of BH spins. 
Assuming that the growth
of supermassive BHs is dominated by their mergers,
\citet{wc95} demonstrated that the excess of radio quiet quasars
could be explained by the rarity of major galaxy
mergers, resulting in a `bottom-heavy' distribution of BH
spins. This is because in such a scenario, fast rotating
BHs can only be produced by mergers of two BHs with
comparable masses, and this in turn implies similar masses of
merging galaxies. However, as  was pointed out by
\citet{ms96a}, the growth of BHs in AGNs is very likely dominated by
accretion processes, which are known to be able to spin up BHs
very efficiently \citep{Bar70}. Moderski \& Sikora argued
nevertheless that the spin paradigm could still be at work, and
showed that the bottom-heavy distribution of the BH spin might be
obtained provided the accretion history of most of AGNs was marked
by many small-mass accretion events with randomly oriented angular
momenta \citep[see also][]{msl98}. Such events, 
due to Bardeen-Petterson effect \citep{bp75},
lead to the formation of co- and counter-rotating disks and, in consequence,
to BH spins fluctuating around zero (with very small amplitudes). Since
during their evolution disk galaxies have avoided major mergers
\citep{hbhe07}, BHs in spiral-hosted AGNs had much larger chance
avoiding spinning-up by massive accretion events than their cousins
in giant ellipticals.

However, these simple versions of the spin paradigm were
challenged, both from the observational and theoretical
perspectives. Observationally, this is because: (i) luminous quasars, 
irrespective on their radio-loudness,
were recently found to be hosted  by giant ellipticals
(see Floyd et al. 2004, and references therein); (ii) several
independent investigations of the AGN accretion radiation
efficiency using the So{\l}tan's type of argument \citep{sol82} 
indicate that the majority of quasars are powered by fast rotating
BHs \citep{yt02,el02,mrg04,wchm06} \citep[but see][]{swm06}; (iii)
low luminosity AGNs (LLAGNs), including local Seyfert galaxies
and LINERs, are found to have higher than previously thought
radio-to-optical nuclear flux ratios, placing them rather in the
category of radio loud objects \citep{hp01,ho02}.
Theoretically, this is because: (iv) the Blandford-Znajek
mechanism was claimed to be not efficient enough to explain jet
energetics in the most radio-loud quasars \citep{ga97}; (v)
the possibility of formation of counter-rotating accretion
disks in AGN systems was questioned \citep{np98,vmqr05}. In
addition, recent investigations of the jet (radio) activity in XRBs
could suggest that the accretion rate is the only parameter
controlling the jet production efficiency in these systems
\citep{gfp03,fbg04}. As a consequence of all the above, the
`main stream' AGN models in the last years became those
with a jet production related to accretion processes 
exclusively \citep{nbb05,kjf06}.

Does this mean that the spin paradigm is `dead'? Not at all. As
was shown and discussed by \citet[][hereafter SSL07]{ssl07}
and \citet{vsl07}, several challenges listed above are not
well justified, while other can be overcome after 
adopting certain modifications to the spin paradigm. In
particular, \citet{klop05} demonstrated analytically and
\citet{lp06} confirmed numerically that the formation of
counter rotating disks in AGNs is possible. Also,
\citet{hk06} and \citet{kinn06ap,kinn06mn} showed (using
different GR MHD simulations) that the efficiency of 
extraction of the BH rotational energy can be much larger than 
indicated by the formula derived originally by \citet{bz77}
using 1st order perturbation method and adopting the
\citet{ss73} disk model. Finally, SSL07 demonstrated
that the host-related radio bimodality  of AGNs {\it remains}
real at {\it all} accretion rates. On the other hand, modifications
of the spin paradigm are indeed required in order to reconcile
the observed radio-bimodality of powerful (elliptical-hosted)
quasars with the requirement that all BHs in quasars have large
spins. Addressing this problem, SSL07 proposed that the fact that
most of the elliptical-hosted quasars are radio quiet results from
the suppression (``quenching") of jet production at high accretion
rates. Such a suppression is in fact directly observed in the
transient XRBs: the observations of micro-quasar GRS 1915-105
indicate for example that two different accretion modes may exist at
high accretion rates, and that only during one of them the efficient
jet production proceeds \citep{fbg04}. In their modified version of the
spin paradigm, SSL07 anticipated such a two accretion-mode scenario
for the elliptical-hosted AGNs accreting at high (Eddington) rates,
assuming in particular that the quasar jets produced by the rotating
supermassive BHs can avoid suppression only if the efficient
collimation by the MHD outflows from the outer parts of the
accretion disk is the case.

This article is based in its large parts on the work
of SSL07, and is organized as follows. In \S2, the dependence of
the AGN radio-loudness on the Eddington ratio is presented; in \S3,
the multi-accretion-event scenarios are investigated; in \S4, 
the challenges to the spin-paradigm are discussed and critically
re-examined; and in \S5 the final conclusions are listed.

\section{Radio-loudness}

In order to quantify the jet production efficiency in AGNs,
\citet{kss89} introduced a `radio-loudness' parameter
defined as the ratio of radio ($5$\,GHz) and optical (B-band)
spectral flux densities, $\mathcal{R} = L_{\nu}(5{\rm
GHz})/L_{\nu}({\rm B})$. Based on the claimed bimodality of
opticaly-selected PG quasars, they established $\mathcal{R}=10$
as a borderline between the radio-loud and radio-quiet
objects. The methods of estimating masses of BH in
galactic nuclei developed in the 1990s \citep[see][and refs.
therein]{wu02} allowed to study the dependence of this
radio-loudness  parameter on the Eddington ratio $\lambda$,
which is defined as the ratio of the bolometric accretion
luminosity to the Eddington luminosity. Such a dependence was first
presented by \citet{ho02} for an AGN sample composed
of low luminosity sources (LLAGNs: weak local
Seyfert galaxies and LINERs), `classical' Seyferts selected
from the Palomar and CfA surveys, and PG quasars. He showed
that the radio-loudness increases with decreasing
Eddington-ratio and that, according to the Kellermann's et al.
definition, practically all the LLAGNs are radio-loud. 
Around the same time, a trend for the increasing
radio-loudness with the decreasing accretion luminosity was
discovered also in the low/hard states of XRBs
\citep{gfp03}. Following these results, \citet{mgf03}
constructed a `fundamental plane' which attempts unifying
the dependence of radio activity on  accretion rate for all the BH
(i.e., galactic and extragalactic) accretion systems.

Such studies have been extended by SSL07 by considering an AGN
sample enlarged  by the addition of radio selected
quasars, broad line radio galaxies (BLRGs) and FR~I radio galaxies.
When compared with the objects considered previously by
\citet{ho02}, the newly included sources form a separate pattern
(though of a similar shape) on the $\mathcal{R}-\lambda$ plane,
being in particular $2-3$ orders radio louder than LLAGNs and
Seyferts with comparable accretion luminosity (see Figures 1 and
2). At the first glance, one might suspect that such a
two-pattern structure is a result of selection effects. However, 
noting that no AGN in the upper,
`radio-loud' branch is hosted by a disk galaxy, one can
conclude that the two revealed upper and lower patterns
represent, at least, the upper bounds for the
radio-loudness of AGNs hosted by elliptical and disk galaxies,
respectively. The relatively complete samples of LLAGNs and
low luminosity radio galaxies analyzed by
\citet{tw03}, \citet{ccm05}, and \citet{pbb07}, indicate that
these are indeed the real distributions of the parameter
$\mathcal{R}$ in the case of low-$\lambda$ objects.

\begin{figure}
\begin{center}
\includegraphics*[width=8cm]{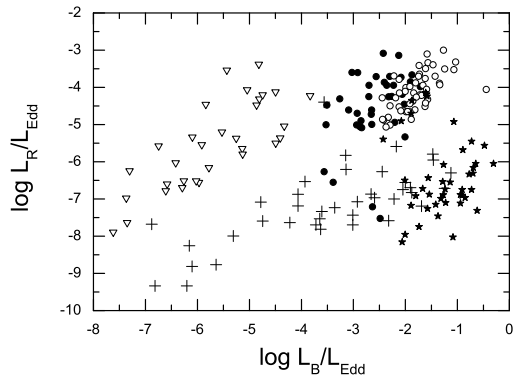}
\end{center}
\caption{Total $5$\,GHz luminosity vs. B-band nuclear luminosity in 
the Eddington units. BLRGs are marked by filled circles, radio-loud quasars 
by open circles, Seyfert galaxies and LINEARs by crosses, FRI radio galaxies 
by open triangles, and PG quasars by filled stars (from SSL07).}
\end{figure}

\begin{figure}
\begin{center}
\includegraphics*[width=8cm]{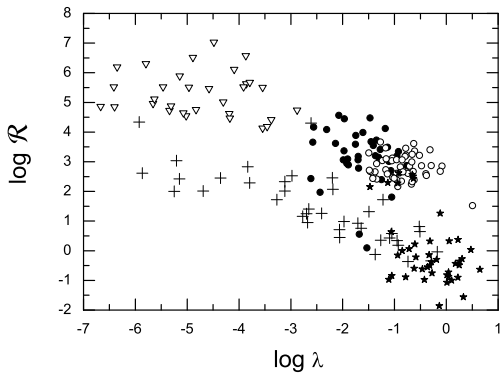}
\end{center}
\caption{Radio-loudness $\mathcal{R}$ vs. Eddington ratio $\lambda$. Different 
types of AGNs are denoted in the same way as in Fig. 1 (from SSL07).}
\end{figure}

The situation changes for the high-accretion rate
(high-$\lambda$) sources, represented by quasars and NLS1 galaxies.
In particular, the radio-loudness distribution of quasars is
very broad and `bottom-heavy', despite that probably most of them are
hosted by giant ellipticals \citep{fkd04}. This cannot be
entirely due to selection effects. Hence, the significant
differences in the jet production efficiency in these objects,
as observed in Figures 1 and 2, seems not to be related to the
morphologies of their hosts. We note, that a bimodal
distribution (instead of a continuous one) of the
radio-loudness in quasar sources was claimed by \citet{kss89},
\citet{mpm90}, and \citet{smw92}, although the most recent studies
on this issue, based on the deep radio and optical surveys, are not
very conclusive \citep{wbg00,imk02,cmcd03,ccmd03,laor03,whb07}.

At the intermediate Eddington-ratios, AGNs hosted by giant
elliptical galaxies and located in the lower (`radio-quiet')
sequence of Figures 1 and 2, are represented in our sample  by
only four objects. However, recent discoveries of many 
elliptical galaxies with very broad Balmer lines and very massive
black holes but very weak radio emission \citep{ssh03,wl04}
strongly indicate that rarity of such objects in the available
AGN catalogs might be due to selection effects. Hence,
it is plausible that also at the intermediate accretion
luminosities most of AGNs hosted by giant ellipticals are
radio-quiet.

SSL07 studied also the dependence of the radio-loudness
parameter on the BH mass. The results are presented in Figure
3. One can see that AGNs with the black hole masses
$\mathcal{M}_{\rm BH} > 10^8 \mathcal{M}_{\odot}$ reach values of
$\mathcal{R}$ three orders of magnitude larger than AGNs with black
hole masses $< 3 \times  10^7 \mathcal{M}_{\odot}$. A relatively
smooth transition between those two populations is caused, most
likely, by the overlap between the sources hosted by disc and
elliptical galaxies. The errors in the black hole mass
estimations can also have a similar effect. It is then
interesting to compare Figure 3 with the analogous figures
restricted to high Eddington-ratio objects, and presented by
\citet{laor00} and \citet{mcj04}. One can see that in all the
cases there is a difference of about $3$ orders of magnitude between
the maximal radio-loudness of AGNs with 
$\mathcal{M}_{\rm BH}/\mathcal{M}_{\odot}> 10^8$, and those with less massive
BHs. However, because in the sample studied by SSL07 the
objects with very low values of the $\lambda$ parameter are included
as well, the boundaries of maximal radio-loudness for lower and higher
mass BHs are now located at much larger values of
$\mathcal{R}$. This effect is a simple consequence of the
increasing radio-loudness with the decreasing
Eddington-ratio. Because of this, the upper boundaries on
Figure 3 are determined by low-$\lambda$ objects, i.e., by
Seyferts and LINERs at low BH masses, and by FR~I radio galaxies at
high values of $\mathcal{M}_{\rm BH}$.

\begin{figure}
\begin{center}
\includegraphics*[width=8cm]{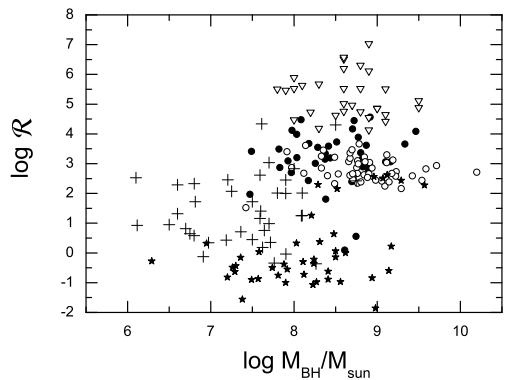}
\end{center}
\caption{Radio loudness $\mathcal{R}$ vs. black hole mass 
$\mathcal{M}_{\rm BH}$ in the solar units.
Different types of AGNs are denoted in the same way as in Fig. 1 (from SSL07).}
\end{figure}

Let us mention in this context the case of nearby galaxies
which show marginal (if any) signatures of the central (accretion)
activity, and for which precise determinations of the profiles of
the stellar distribution are available. In particular, the most
recent studies led to the discovery of a strong
correlation between the type of the surface brightness
distribution in the nuclear parts of the galaxy and its radio
luminosity. Namely, the `core-galaxies' -- i.e., the ones with
shallow central stellar density distributions -- are much
radio-louder than the power-law/cuspy galaxies \citep{cb06,cb07}. It
is then interesting to note that the core-galaxies are on
average much more massive than the power-law galaxies,
being usually identified as giant ellipticals, while the
power-law stralight profiles are common in disk galaxies or
disky-ellipticals \citep{lgf07,cb07}.

\section{The spin paradigm}

SSL07 investigated the possibility that the parameter
responsible for the host-morphology related split of the upper
radio-bounds of AGNs in the $\mathcal{R}-\lambda$ plane, as
discussed above, is the spin of a central black hole, $a
\equiv J/J_{\rm max}$, where $J$ is the angular momentum of the
black hole and $J_{\rm max}= G \, \mathcal{M}_{\rm BH}^2/c$. The
conditions required for that are:
\begin{enumerate}
\item BHs in disk galaxies should avoid efficient spinning-up by 
the accretion disks;
\item it should be possible to reconcile the spin paradigm with the mismatch 
between the spin and
radio-loudness distributions of the high-$\lambda$ AGNs hosted by elliptical 
galaxies.
\end{enumerate}
As  was demonstrated by
\citet{ms96a}, BHs can avoid too intensive spinning-up if their
evolution is composed of many small-mass
accretion events with randomly oriented angular momenta. In
such a case, provided that the event masses are small enough to
avoid an alignment of the BH with distant portions of
the accretion disk \citep{rees78}, the accretion proceeds via
a comparable number of co- and counter-rotating disks. As a
result, the BH spin undergoes fluctuations around the zero
value with a very low amplitude, or quickly drops to
small values if starting from a large one (see Figures 4 \& 5).

\begin{figure}
\begin{center}
\includegraphics*[width=8cm]{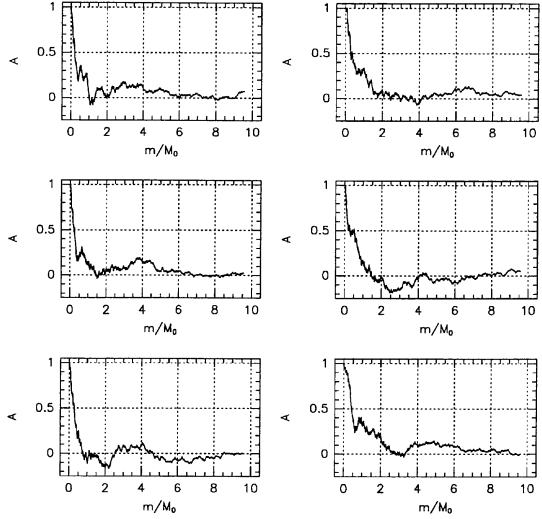}
\end{center}
\caption{The spin evolution of a black hole for the initial spin $A_0=1$ 
and the accretion event masses 
$\Delta m=0.01 \mathcal{M}_{\rm BH, \, 0}$ \citep[from][]{ms96a}.}
\end{figure}

\begin{figure}
\begin{center}
\includegraphics*[width=8cm]{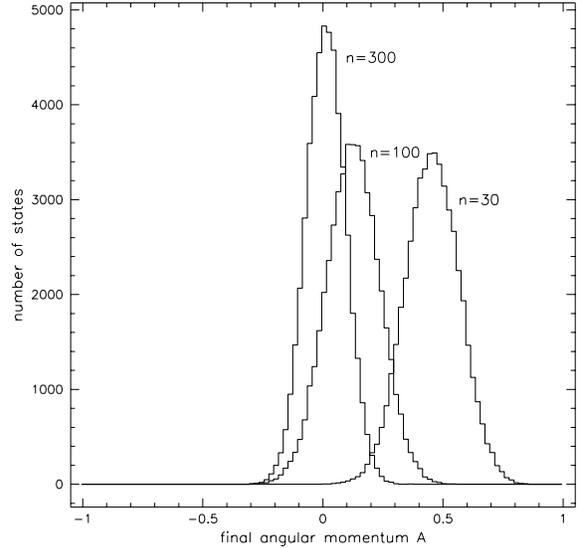}
\end{center}
\caption{Histogram of the final states of 50000 evolutionary tracks computed 
for $A_0=1$ and $\Delta m= 0.01 \mathcal{M}_{\rm BH,\,0}$. The number of 
accretion events is marked on the curves \citep[from][]{msl98}.}
\end{figure}

Quantitatively, the condition to avoid the BH alignment during an accretion 
event of a mass $m$ is
\begin{equation}
m \ll m_{\rm align} \equiv {a \over \sqrt{r_{\rm w}}} \, \mathcal{M}_{\rm BH} \, ,
\end{equation}
\noindent
where $r_{\rm w} \equiv R_{\rm w}/ R_{\rm S}$, 
$R_{\rm w}$ is the Bardeen-Petterson warp radius, i.e. the distance at 
which the accretion time scale is equal to the time scale of 
the Lense-Thirring precession \citep{Wilk72}, and
$R_{\rm S} \equiv 2G\,\mathcal{M}_{\rm BH}/c^2$ is the Schwarzschild radius. 
For the \citet{ss73} disk model, for example, the warp is formed in 
the `middle region of the disk', namely at
\begin{equation}
r_{\rm w} = 3.6 \times 10^3 \, a^{5/8} \mathcal{M}_8^{1/8} f_{\rm Edd}^{-1/4} \alpha^{-1/2} \left({\nu_1 \over \nu_2}\right)^{5/8}\!\!,
\end{equation}
where $\mathcal{M}_8 \equiv \mathcal{M}_{\rm BH}/ 10^8 \, \mathcal{M}_{\odot}$, 
$f_{\rm Edd} = \dot{m}\,c^2/ L_{\rm Edd}$ is the accretion rate expressed in 
the Eddington units 
$L_{\rm Edd} \equiv 4 \pi \, G \mathcal{M}_{\rm BH} \, 
m_{\rm p} c / \sigma_{\rm T}$, $\alpha$ is the Shakura-Sunyaev parameter, 
$\nu_1$ is the viscosity related to the `planar' shear within the disk, and 
$\nu_1$ is the viscosity related to the `vertical' shear. According to 
\citet{pp83}, $1 < \nu_2/\nu_1 < 1/(2\alpha^2$). With this, the alignment mass 
reads as
\begin{equation}
{m_{\rm align} \over \mathcal{M}_{\rm BH}} \simeq 10^{-4} \, {a^{3/8} f_{\rm Edd}^{1/4} \over \mathcal{M}_8^{1/8}} \, \left({\alpha \over 0.1}\right)^{1/2} \left({\nu_1 \over \nu_2}\right)^{-5/8}\!\!.
\end{equation}
Such an event-mass limit is severe but, on the other hand,
consistent with the observations indicating very short lifetimes of
individual accretion events in Seyfert galaxies \citep{cam99,kob06},
and revealing random orientations of jets relative to the host
galaxy axis \citep{ksc00,sau01}. It should be also noted that
fuelling of AGNs in disk galaxies is presumably not related to the
galaxy mergers, and can be provided by molecular clouds
\citep{hh06}. Alternatively, low values of BH spins in such
systems could be assured if the BH growth thereby is
dominated by mergers with intermediate mass BHs -- relics of
Population III stars, or BHs formed in young stellar clusters
\citep[see][and references therein]{mfr06}.

In contrast to spiral galaxies, giant ellipticals underwent at least
one major merger in the past \citep[see, e.g.,][]{hbhe07}. Such
mergers are followed by accretion events which involve too much mass
to satisfy the condition given by Equation 3. Then, regardless of
whether the accretion disk was initially counter- or
co-rotating, all the disks end up as co-rotating due to the
alignment process. Provided that $m >> m_{\rm align}$, such disks
spin-up black holes to large values of $a$, very likely up to
even $a > 0.95$ if only $m \sim M_{\rm BH}$ \citep{ms96b}. This
scenario is in agreement with the large average spin of 
BHs in quasars as inferred from the comparison between the
local BH mass density and the amount of radiation produced by
luminous quasars \citep{sol82,yt02,el02,mrg04,wchm06}. However,
since the majority of quasars is radio-quiet, the jet production
in most of them should be suppressed. Such a bimodality of a jet
suppression at high accretion rates can be connected with a
bimodality of a jet collimation, and this, in turn, with a
bimodality of the accretion states. The collimation of a
central relativistic outflow can be provided by a
non-relativistic MHD wind/jet produced by  
an accretion disk. Such a double jet structure was originally
proposed by \citet{spa89}, and was investigated further by
\citet{bt05}, \citet{gtb05}, and \citet{bn06}. The quasar
systems with such efficiently collimated jets can be rare indeed due
to the difficulties in developing the required large-scale poloidal
magnetic fields in the geometrically thin accretion disks. Such
fields may, however, develop stochastically, leading to an
intermittent jet production \citep{lpk03,mp06}. Alternatively,
large-scale poloidal fields may be carried to the inner portions of
the disks from larger distances by the drifting isolated patches of
the accreting matter \citep{su05}.

\section{Discussion}

Recent studies of the dependence of the
radio-loudness on the Eddington-ratio in AGNs and XRBs indicate that
the jet-to-accretion power ratio increases with the decreasing
accretion rate, and that at the high accretion rates the jet
production is strongly suppressed in majority of objects
\citep{ho02,gfp03,mgf03}. In addition, the previous claims
that powerful extragalactic radio sources avoid disk galaxies
\citep{mms64,xlb99} have been confirmed, and, as shown by
SSL07, such a radio bimodality persists over an entire 
investigated range of the Eddington-ratios.

Assuming that jets are powered by rotating BHs, the
galaxy-morphology related radio-bimodality can be explained if BH
spins in giant ellipticals are, on average, much larger than
in disk galaxies. The best and the most direct way to
verify this conjecture is to measure BH spins in different types of
galaxies. First attempts are already undertaken using profiles of
the fluorescent iron X-ray line emitted in some AGNs by
the innermost portions of the accretion disks
\citep[e.g.][]{laor91,bd04,fab07}. Unfortunately, the quality
of the available data is too poor to disentangle 
the effects connected with the BH spin from the effects
of a warm absorption \citep{rpt04,nog06,dg06}, and/or the
effects of a superposition of differently shaped spectra 
produced in different flux states \citep{mtr07}. We would like to
point out that MCG-6-30-15, considered often to be the best
evidenced case showing the fast rotating BH in a Seyfert galaxy, is
hosted in fact by the E/S0 galaxy \citep{fwm00}. 
Also, the Eddington ratio in this system is very high, $\lambda
\simeq 0.4$ \citep{mgug05}, and so the jet production thereby is
likely to be suppressed, even if MCG-6-30-15 hosts indeed the fast
rotating BH.

SSL07 and \citet{vsl07} investigated possible evolutionary scenarios
which may lead to the galaxy-morphology related BH spin
bimodality of AGNs. Assuming that the growth of supermassive
BHs is dominated by the accretion, they showed that, in
order to keep small BH spins in the multi-accretion event scenario
\citep{ms96a}, the event masses must be smaller than $10^{-4}
\, \mathcal{M}_{\odot}$. Such small-mass accretion events can
be provided by stochastically captured molecular clouds
\citep{hh06}. These may form from a cold gas
streaming/dropping onto the galaxy from cosmological
filaments. Such an inflow is predicted to take place in galaxies
with the dark matter halo masses smaller than $10^{12} \,
\mathcal{M}_{\odot}$ \citep{db06}, and is supposed to protect
the disk galaxies against destruction by the multiple minor
mergers \citep{bjc07}.

The modified spin paradigm described above is consistent
with the recent finding that the radio-loudness correlates with
stellar brightness profiles in the nuclear portions of active
galaxies \citep{cb06,cb07}. Namely, the inner regions of radio-loud
galaxies display star deficient cores. Such cores, in turn,
reside preferentially in giant ellipticals \citep[see,
e.g.,][]{lgf07}. On the other hand, radio-quiet galaxies,
including nearby low-luminosity Seyferts, display instead
cuspy brightness profiles. And these are found preferentially
in disk galaxies. Hence, noting that the core stellar nuclei
result from merging BHs following galaxy mergers
\citep{emo91,mm01,rhf02,vmh03}, Balmaverde \& Capetti's discovery
supports our conjecture that the galaxy-morphology related
spin bimodality results from the different evolutionary tracks
of BHs in disk and elliptical galaxies: in the former case
being dominated by randomly oriented small-mass accretion events, in
the latter case by massive accretion events which follow galaxy
mergers.

\section{Conclusions}

In summary, we conclude that:
\bigskip

\begin{enumerate}
\item The maximal values of the radio-loudness parameter of AGNs hosted by 
giant elliptical galaxies are by $\sim 3$ orders of magnitude larger than 
of AGNs hosted by disc galaxies.
\item Both populations of spiral-hosted and elliptical-hosted AGNs show 
a similar dependence of the upper bounds of the radio-loudness parameter on 
the Eddington ratio; the radio-loudness increases with the decreasing 
Eddington ratio, faster at the high accretion rates, and slower at the low 
accretion rates.
\item The very large, host-morphology related difference between 
the radio-loudness 
reachable by AGNs in disc and elliptical galaxies can be explained by 
the scenario according to which
\begin{itemize}
\item the spin of a black hole determines the jet power;
\item central black holes can reach large spins only in early type galaxies 
(following major mergers), and not (in a statistical sense) in spiral galaxies.
\end{itemize}
\item The broad, `bottom-heavy' distribution of the radio-loudness in quasars 
is not related to the distribution of the BH spin; however, it is still 
the BH spin which mediates launching of a jet and determines the upper bound 
of the radio-loudness, whereas the (intermittent) suppression of a jet 
production can be connected with the absence of the jet collimation by 
an MHD wind from an accretion disk.
\end{enumerate}

\section*{Acknowledgments} {\L}.S. acknowledges support by the MEiN
grant 1-P03D-00329. M.S. thanks KIPAC/SLAC for the hospitality 
during the completion of this manuscript. 

\bibliographystyle{elsart-harv}

\begin{thebibliography}{}

\bibitem[Bardeen(1970)]{Bar70}
Bardeen, J.M. 1970, Nature, 226, 64

\bibitem[Bardeen \& Petterson(1975)]{bp75}
Bardeen, J.M., \& Petterson, J.A. 1975, ApJ, 195, L65

\bibitem[Beckwith \& Done(2004)]{bd04}
Beckwith, K., \& Done, C. 2004, MNRAS, 352, 353

\bibitem[Beskin \& Nokhrina(2006)]{bn06}
Beskin, V.S., \& Nokhrina, E.E. 2006, MNRAS, 367, 375

\bibitem[Blandford(1990)]{bland90}
Blandford, R.D. 1990, in Active Galactice Nuclei,
ed. T.J.-L. Courvoisier \& M. Mayor (Saas-Fee Advanced Course 20; Berlin:Springer), 161

\bibitem[Blandford \& Znajek(1977)]{bz77}
Blandford, R.D., \& Znajek, R.L. 1977, MNRAS, 179, 433

\bibitem[Bogovalov \& Tsinganos(2005)]{bt05}
Bogovalov, S., \& Tsinganos, K. 2005, MNRAS, 357, 918

\bibitem[Bournaud, Jog, \& Combes(2007)]{bjc07}
Bournaud, F., Jog, C.J., \& Combes, F. 2007, ApJ, 670, 237

\bibitem[Capetti et al.(1999)]{cam99}
Capetti, A., Axon, D.J., Macchetto, F.D., et al. 1999, ApJ, 516, 187

\bibitem[Capetti \& Balmaverde(2006)]{cb06}
Capetti, A., \& Balmaverde, B. 2006, A\&A, 453, 27

\bibitem[Capetti \& Balmaverde(2007)]{cb07}
Capetti, A., \& Balmaverde, B. 2007, A\&A, 469, 75

\bibitem[Chiaberge et al.(2005)]{ccm05}
Chiaberge, M., Capetti, A., \& Macchetto, F.D. 2005, ApJ, 625, 716

\bibitem[Cirasuolo et al.(2003a)]{cmcd03}
Cirasuolo, M., Magliocchrtti, M., Celotti, A., \& Danese, L. 2003,
MNRAS, 341, 993

\bibitem[Cirasuolo et al.(2003b)]{ccmd03}
Cirasuolo, M., Celotti, A., Magliocchrtti, M., \& Danese, L. 2003,
MNRAS, 346, 447

\bibitem[Dekel \& Birnboim(2006)]{db06}
Dekel, A. \& Birnboim, Y. 2006, MNRAS, 368, 2

\bibitem[Done \& Gierli\'nski(2006)]{dg06}
Done, C., \& Gierli\'nski, M. 2006, MNRAS, 367, 659

\bibitem[Ebisuzaki et al.(1991)]{emo91}
Ebisuzaki, T., Makino, J., Okumura, S.K. 1991, Nature, 354, 212

\bibitem[Elvis et al.(2002)]{el02}
Elvis, M., Risaliti, G., \& Zamorani, G., 2002, ApJ, 565, L75

\bibitem[Fabian(2007)]{fab07}
Fabian, A.C. 2007, arXiv0711.2976

\bibitem[Fender et al.(2004)]{fbg04}
Fender, R.P., Belloni, T.M., \& Gallo, E. 2004, MNRAS, 355, 1105

\bibitem[Feruuit et al.(2000)]{fwm00}
Ferruit, P., Wilson, A.~S., \& Mulchaey, J.\ 2000, ApJS, 128, 139

\bibitem[Floyd et al.(2004)]{fkd04}
Floyd, D.J.E., Kukula, M.J., Dunlop, J.S., et al. 2004, MNRAS, 355, 196

\bibitem[Gallo et al.(2003)]{gfp03}
Gallo, E., Fender, R.P., \& Pooley, G.G. 2003, MNRAS, 344, 60

\bibitem[Ghosh \& Abramowicz(1997)]{ga97}
Ghosh, P. \& Abramowicz, M.A. 1997, MNRAS, 292, 887

\bibitem[Gracia et al.(2005)]{gtb05}
Gracia, J., Tsinganos, K., \& Bogovalov, S.V. 2005, A\&A, 442, L7

\bibitem[Hawley \& Krolik(2006)]{hk06}
Hawley, J.F., \& Krolik, J.H. 2006, ApJ, 641, 103

\bibitem[Ho(2002)]{ho02}
Ho, L.C. 2002, ApJ, 564, 120

\bibitem[Ho \& Peng(2001)]{hp01}
Ho, L.C., \& Peng, C.Y. 2001, ApJ, 555, 650

\bibitem[Hopkins et al.(2007)]{hbhe07}
Hopkins, P.F., Bundy, K., Hernquist, L., \& Ellis,~R.S. 2006,
ApJ, 659, 976

\bibitem[Hopkins \& Hernquist(2006)]{hh06}
Hopkins, P.F., \& Hernquist, L. 2006, ApJS, 166, 1

\bibitem[Ivezi\'c et al.(2002)]{imk02}
Ivezi\'c, \u Z, Menou, K., Knapp, G.R., et al. 2002, AJ, 124, 2364

\bibitem[Kellermann et al.(1989)]{kss89}
Kellermann, K.I., Sramek, R., Schmidt, M., et al. 1989, AJ, 98, 1195

\bibitem[Kharb et al.(2006)]{kob06}
Kharb, P., O'Dea, C.P., Baum, S.A., et al. 2006, ApJ, 652, 177

\bibitem[King et al.(2005)]{klop05}
King, A.R., Lubow, S.H., Ogilvie, G.I., \& Pringle,~J.E. 2005,
MNRAS, 363, 49

\bibitem[Kinney et al.(2000)]{ksc00}
Kinney, A.L., Schmitt, H.R., Clarke, C.J., et al. 2000, ApJ, 537, 152

\bibitem[K\"ording et al.(2006)]{kjf06}
K\"ording, E.G., Jester, S., \& Fender, R. 2006, MNRAS, 372, 1366

\bibitem[Laor(1991)]{laor91}
Laor, A. 1991, ApJ, 376, 90

\bibitem[Laor(2000)]{laor00}
Laor, A., 2000, ApJ, 543, L111

\bibitem[Laor(2003)]{laor03}
Laor, A., 2003, ApJ, 590, 86

\bibitem[Lauer et al.(2007)]{lgf07}
Lauer, T.R., Gebhardt, K., Faber, S.M., et al. 2007, ApJ, 664, 226

\bibitem[Livio et al.(2003)]{lpk03}
Livio, M., Pringle, J.E., \& King, A.R. 2003, ApJ, 593, 184

\bibitem[Lodato \& Pringle(2006)]{lp06}
Lodato, G., \& Pringle, J.E. 2006, MNRAS, 368, 1196

\bibitem[Maccarone et al.(2003)]{mgf03}
Maccarone, T.J., Gallo, E., \& Fender, R. 2003, MNRAS, 345, L19

\bibitem[Malkan(1984)]{Mal84}
Malkan, M. 1984, ApJ, 287, 555

\bibitem[Mapelli et al.(2006)]{mfr06}
Mapelli, M., Ferrara, A., \& Rea, N. 2006, MNRAS, 368, 1340

\bibitem[Marconi et al.(2004)]{mrg04}
Marconi, A., Risaliti, G., Gilli, R., et al. 2004, MNRAS, 351, 169

\bibitem[Matthews et al.(1964)]{mms64}
Matthews, T.A., Morgan, W.W., \& Schmidt, M. 1964, ApJ, 140, 35

\bibitem[Mayer \& Pringle(2006)]{mp06}
Mayer, M., \& Pringle, J.E. 2006, MNRAS, 368, 379

\bibitem[McHardy et al.(2005)]{mgug05}
McHardy, I.~M., Gunn,
K.~F., Uttley, P., \& Goad,~M.R.\ 2005, MNRAS, 359, 1469

\bibitem[McKinney(2006a)]{kinn06ap}
McKinney, J.C. 2005, ApJ, 630, L5

\bibitem[McKinney(2006b)]{kinn06mn}
McKinney, J.C. 2006, MNRAS, 368, 1561

\bibitem[McLure \& Jarvis(2004)]{mcj04}
McLure, R.J., \& Jarvis, M.J. 2004, MNRAS, 353, L45

\bibitem[Miller et al.(1990)]{mpm90}
Miller, L., Peacock, J.A., \& Mead, A.R.G. 1990, MNRAS, 244, 207

\bibitem[Miller et al.(2007)]{mtr07}
Miller, L., Turner, J., \& Reeves,~J. 2007, poster presentation at
the conference ``The Suzaku X-ray Universe'', San Diego, 10-12 Dec. 2007

\bibitem[Milosavljevic \& Merritt(2001)]{mm01}
Milosavljevic, M., \& Merritt, D. 2001, ApJ, 563, 34

\bibitem[Moderski \& Sikora(1996a)]{ms96a}
Moderski, R., \& Sikora, M. 1996a, A\&AS, 120C, 591

\bibitem[Moderski \& Sikora(1996b)]{ms96b}
Moderski, R., \& Sikora, M. 1996b, MNRAS, 283, 854

\bibitem[Moderski et al.(1998)]{msl98}
Moderski, R., Sikora, M., \& Lasota, J.-P. 1998, MNRAS, 301, 142

\bibitem[Nandra et al.(2006)]{nog06}
Nandra, K., O'Neill, P.M., George, I.M., Reeves, U.J.N., \& Turner, T.J.
2006, Astr. Nach. 327, 1039

\bibitem[Natarajan \& Pringle(1998)]{np98}
Natarajan, P., \& Pringle, J.E. 1998, ApJ, 506, L97

\bibitem[Nipoti et al.(2005)]{nbb05}
Nipoti, C., Blundell, K.M., \& Binney, J. 2005, MNRAS, 361, 633

\bibitem[Panessa et al.(2007)]{pbb07}
Panessa, F., Barcons, X., Bassani, L., et al. 2007, A\&A, 467, 519

\bibitem[Papaloizou \& Pringle(1983)]{pp83}
Papaloizou, J.C.B., \& Pringle, J.E. 1983, MNRAS, 202, 1181

\bibitem[Ravindranath et al.(2002)]{rhf02}
Ravindranath, S., Ho, L.C., \& Fillipenko, A.V. 2002, ApJ, 566, 801

\bibitem[Rees(1978)]{rees78}
Rees, M.J. 1978, Nature, 275, 516

\bibitem[Reeves et al.(2004)]{rpt04}
Reeves, J.N., Porquet, D., \& Turner, T.J. 2004, ApJ, 615, 150

\bibitem[Sandage(1965)]{Sandage}
Sandage, A. 1965, ApJ, 141, 1560

\bibitem[Schmitt et al.(2001)]{sau01}
Schmitt, H.R., Antonucci, R.R.J.,
Ulvestad, J.S., et al. 2001, ApJ, 555, 663

\bibitem[Shakura \& Sunyaev(1973)]{ss73}
Shakura, N.I., \& Sunyaev, R.A. 1973, A\&A, 24, 337

\bibitem[Shankar et al.(2007)]{swm06}
Shankar, F., Weinberg, D.H., \& Miralda-Escud\'e, J. 2007,
arXiv:0710.4488

\bibitem[Sikora et al.(2007)]{ssl07}
Sikora, M., Stawarz, {\L}, \& Lasota, J.-P. 2007, 658, 815 (SSL07)

\bibitem[Smith et al.(1986)]{shbrb86}
Smith, E.P., Heckman, T.M., Bothun, G.D., Romanishin, W., \& Balick, B.
1986, ApJ, 306, 64

\bibitem[Sol et al.(1989)]{spa89}
Sol, H., Pelletier, G., \& Asseo, E. 1989, MNRAS, 237, 411

\bibitem[So{\l}tan(1982)]{sol82}
So{\l}tan, A. 1982, MNRAS, 200, 115

\bibitem[Spruit \& Uzdensky(2005)]{su05}
Spruit, H.C., \& Uzdensky, D.A. 2005, ApJ, 629, 960

\bibitem[Stocke et al.(1992)]{smw92}
Stocke, J.T., Morris, S.L., Weymann, R.J., et al. 1992, ApJ, 396, 487

\bibitem[Strateva et al.(2003)]{ssh03}
Strateva, I.V., Strauss, M.A., Hao, L., et al. 2003, AJ, 126, 1720

\bibitem[Strittmatter et al.(1980)]{shp80}
Strittmatter, P.A., Hill, P., Pauliny-Toth, , I.I.K., et al. 1980,
A\&A, 88, L12

\bibitem[Terashima \& Wilson(2003)]{tw03}
Terashima, Y, \& Wilson, A.S. 2003, ApJ, 583, 145

\bibitem[Volonteri et al.(2003)]{vmh03}
Volonteri, M., Madau, P., Haardt, F. 2203, ApJ, 593, 661

\bibitem[Volonteri et al.(2005)]{vmqr05}
Volonteri, M., Madau, P., Quataert, E., \& Rees, M.J. 2005,
ApJ, 620, 69

\bibitem[Volonteri et al.(2007)]{vsl07}
Volonteri, M., Sikora, M., \& Lasota, J.-P. 2007, ApJ, 667, 704

\bibitem[Wang et al.(2006)]{wchm06}
Wang, J.-M., Chen, Y.-M., Ho, L.C., \& McLure, R.J. 2006, ApJ, 642, L111

\bibitem[White et al.(2000)]{wbg00}
White, R.L., Becker, R.H., Gregg, M.D., et al. 2000, ApJS, 126, 133

\bibitem[White et al.(2007)]{whb07}
White, R.,L., Helfand, D.J., Becker, R.H., Glikman, E., \& de Vries, W.
2007, ApJ, 654, 99

\bibitem[Wilson \& Colbert(1995)]{wc95}
Wilson, A.S., \& Colbert, E.J.M. 1995, ApJ, 438, 62

\bibitem[Woo \& Urry(2002)]{wu02}
Woo, J.-H., \& Urry, C.M. 2002, ApJ, 579, 530

\bibitem[Xu et al.(1999)]{xlb99}
Xu, C., Livio, M., Baum, S. 1999, AJ, 118, 1169

\bibitem[Wilkins(1972)]{Wilk72}
Wilkins, D.C. 1972, Phys. Rev. D., 5, 814

\bibitem[Wu \& Liu(2004)]{wl04}
Wu, X.-B., \& Liu, F.K. 2004, ApJ, 614, 91

\bibitem[Yu \& Tremaine(2002)]{yt02}
Yu, Q., \& Tremaine, S. 2002, MNRAS, 335, 965

\end{thebibliography}

\end{document}